\documentclass[a4paper,twocolumn,showkeys,floatfix,aps,pre,reprint,groupedaddress]{revtex4-1}
\usepackage{amsmath,amssymb}
\usepackage{graphics,graphicx}
\usepackage{dcolumn,bm}
\usepackage{psfrag}
\usepackage{enumerate}

\begin{document}

\title{Avalanche dynamics in hierarchical fiber bundles}

\author{Soumyajyoti Biswas${}^{1}$}
\email{soumyajyoti.biswas@fau.de}
\author{Michael Zaiser${}^{1}$}
\email{michael.zaiser@fau.de}
\affiliation{
${}^1$ WW8-Materials Simulation, Department of Materials Science, Friedrich-Alexander-Universit\"{a}t Erlangen-N\"{u}rnberg, Dr.-Mack-Str. 77, 90762 F\"{u}rth, Germany
}

\date{\today}

\begin{abstract}
Heterogeneous materials are often organized in a hierarchical manner, where a basic unit is repeated over multiple scales.
The structure then acquires a self-similar pattern. Examples of such structure are found in various biological and synthetic materials. The hierarchical structure can have significant consequences for the failure strength and the mechanical response
of such systems. Here we consider a fiber bundle model with hierarchical structure and study the effect of the self-similar arrangement on the avalanche dynamics exhibited by the model during the approach to failure. We show that the failure strength of the model generally decreases in a hierarchical structure, as opposed to the situation where no such hierarchy exists. However, we also report a special arrangement of the hierarchy for which the failure threshold could be substantially above that of a non hierarchical reference structure.  
\end{abstract}


\maketitle


\section{Introduction}
A wide variety of materials, both biological \cite{gau11,smith99,pugno10} and synthetic \cite{dalton03} exhibit structural self-similarity. Essentially, a basic structural unit is repeated multiple times, bridging across length scales, for example, from a single carbon nanotube or molecules of micro fibrils to fibrils, fibril bundles and ropes to the design of a space-elevator cable \cite{pugno08}. The smallest units (nano tubes or micro fibrils) are coupled together to form one unit of the higher level
and so on, eventually reaching a macroscopic scale \cite{gau11}. 

The advantage of a hierarchical structure is its ability to carry load due to the inhibitory effect of the hierarchical structure towards damage propagation. It has been observed that a hierarchical structure prevents crack growth \cite{scirep} and thereby increases the flaw tolerance of the structure compared to a non-hierarchical counterpart of similar system size. Furthermore, the extreme nature of failure statistics, of a chain of fibers for example, tend towards self-averaging behavior, leading to a more predictable response of the system.

While the failure strengths of hierarchical structures have received considerable attention recently \cite{pugno12}, less is known about the intermittent dynamics of such systems in the run-up to failure. Particularly, local failure within one unit of the hierarchy (damage nucleation) leads to re-distribution of load, and this re-distribution and the resulting diffusion of damage are strongly affected by the hierarchical organization of the system. The interplay of the competing effects of local damage nucleation and long-range load re-distribution leading to damage diffusion is manifested in the response statistics of the system, i.e. the avalanche dynamics.

Here we consider a fiber bundle model that is arranged in a hierarchical manner and study the interplay of damage nucleation and load re-distribution in the avalanche response. A fiber bundle model is a generic model for studying response of driven disordered systems leading to fracture \cite{fbm_rmp,fbm_book}. Each  fiber has a finite failure threshold drawn from a probability distribution and therefore can sustain a particular amount of load. Following the failure of a fiber, the remaining fibers in the bundle share its load, which might lead to further failure and so on. The collective response of the model mimics several features manifested in breakdown experiments in disordered solids \cite{wiley_book}. In this particular study, we are interested in the avalanche statistics of the fiber bundle model when the elements/fibers are arranged in a hierarchical structure. We show that the effects of damage nucleation and hierarchically structured load re-distribution and damage diffusion are clearly manifested in the avalanche dynamics. Furthermore, we propose a specific hierarchical structure, which increases the total strength of the system substantially from the usual random placing of the fibers. 

The paper is organized as follows: First we describe the hierarchical fiber bundle model. Then we discuss the effects of various levels of hierarchy on the avalanche statistics of the model and also note the effect of the hierarchical structure on overall strength of the system. Then we propose the structure of hierarchy that maximizes the strength of the overall system, for a given distribution of the failure thresholds of the elementary fibers. Finally, we discuss and summarize the results.

\section{Model}  
A fiber bundle model, in its simplest form, is a series of fibers fixed between two rigid parallel plates and the plates are
pulled apart by a force (load) leading to the eventual collapse of the system. The individual fibers are considered to be linear elastic and perfectly brittle, i.e. their stress-strain curves are linear up to the irreversible failure point where the fibers cease to support any load. The particular points of failures, however, are not the same for all the fibers but are drawn from a distribution function. Here we consider mainly a uniform distribution between $(0:1)$ unless otherwise mentioned. But similarly well-behaved distributions give similar statistics for the failure dynamics. With the application of a load, the weaker fibers break and the total load is carried by the stronger fibers. This increases the load per fiber acting on the remaining fibers and can lead to further breaking events. The system can eventually reach a stable state, where all fibers are stronger than their respective load per fiber, or it can suffer a catastrophic breakdown if no such stable state exists for the applied load. 

The particular load at which stable states cease to exist for a given system of fibers, is the critical load for that system. The behavior of the fiber bundle model near its critical load is well studied. A gradual increase of the load from zero to the critical load, visits all the stable configurations of the model. The fiber breaking between any two successive stable configurations constitute one avalanche, the number of fibers broken in an avalanche defines its size. The size distribution of all the avalanches, from zero to the critical load, is a power-law function. This is similar to what is seen experimentally in disordered systems under time dependent loading (see, for example, ref. \cite{baro}). 

In its simplest form described above, the avalanche statistics of a fiber bundle model is universal with respect to a broad class of failure threshold distributions. The universality, in this case, is guaranteed by the mean-field nature of the model, i.e., following the failure of one fiber, its load is shared equally among the remaining fibers, irrespective of their distance from the failed fiber. A departure from the equal-load-sharing (ELS) mechanism can significantly change the response of the system \cite{fbm_book}. 

Here we consider a hierarchical structure of the model in which the load redistribution is drastically modified from the ELS version. As shown in Fig.1, we consider a division of the system into hierarchically organized modules such that, at the $i$-th level of hierarchy, each one of the $N_i$ fibers in the $i$-th level module represents a module of level $i-1$ consisting of a bundle of $N_{i-1}$ fibers, and so on. For a total of $k$ hierarchy levels, the total number of lowest-level fibers is $N=\prod\limits_{i=1}^k N_i$.

\begin{figure}[tbh]
\centering
\includegraphics[width=8cm, keepaspectratio]{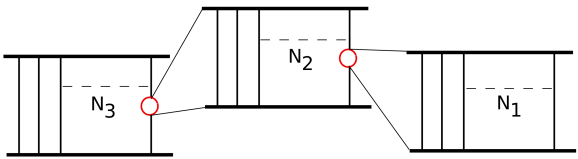} 
\caption{Schematic illustration of a hierarchical fiber bundle.}
\label{schematic}
\end{figure} 

If a fiber in a $i$th level module fails, its load is redistributed equally among all the surviving fibers within the same hierarchical module. If there are no surviving fibers in that module, the module is considered as failed and its load is redistributed among all the surviving level-$(i+1)$ fibers  in the next-higher ($i+1$st) level module. Note that, in this case, while the load increments received by the individual level-$(i+1)$ fibers are equal, due to the possible variations in the number of surviving level-$i$ fibers within each of those units, the load shares for the level-$i$ fibers belonging to different units might differ. 

It is immediately clear that the dynamics is very different from the ELS dynamics described before. Particularly, there is a tendency of accumulation of damage within one unit, until that unit completely fails. This leads to an interesting competition between damage accumulation and load re-distribution. When the unit sizes are very small, damage spreads rapidly within a unit because the load of a failed fiber is redistributed very locally. But as the unit size is small, the whole unit breaks quickly, redistributing the load to other units and thereby diffusing the load concentration. On the other hand, if the unit sizes are large, then there is not much damage accumulation to begin with as the individual units behave 
like ELS fiber bundles. In between these limits one finds a critical size for the units, at which the damage accumulation is enhanced because of load re-distribution confinement and the hierarchical structure becomes weakest. This effect is also visible in the avalanche statistics. 

Below we investigate the dynamics of such hierarchical fiber bundles. We start with the simplest case of two hierarchy levels and later generalize to three and more levels. At the end we revisit the two-level system but with variable sizes for each unit and show how the overall strength could be increased following a particular prescription for the hierarchical structure which correlates unit size with strength of the constituent sub-units. 

\section{Results}
\subsection{Two level hierarchy}

\begin{figure}[tbh]
\centering
\includegraphics[width=8cm, keepaspectratio]{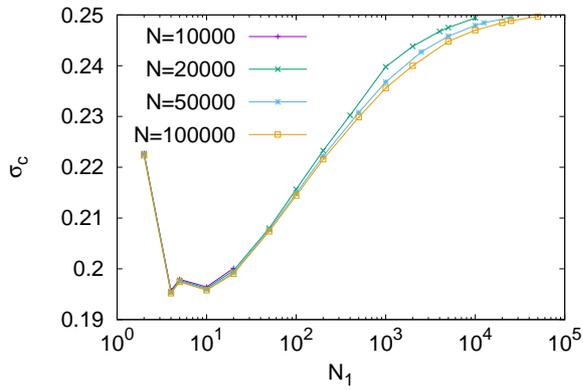} 
\caption{The critical load of a two-level hierarchical fiber bundle model for fixed system size as a function of the level-one unit size. In the two limiting cases, when the unit size is 1 or equal to the full system size, the critical load is trivially $1/4$, which is the result for an ELS fiber bundle with uniform threshold distribution. In all intermediate cases, the critical load is lower due to confinement-enhanced damage accumulation.}
\label{level2_load}
\end{figure}

In its simplest form, we study a two-level hierarchical fiber bundle model. Here we consider a set of $N_2$ fibers, each of which is made up of  $N_1$ fibers, giving the total number of fibers to be $N=N_1N_2$. As mentioned above, if $N$ is kept fixed, the various combinations of $N_1$ and $N_2$ will give rise to competing effects of damage nucleation and damage diffusion. In Fig. \ref{level2_load}, we show the critical load of the model, as a function of $N_1$, for a fixed total number $N$ of elementary fibers. The limits $N_1=1, N_2=N$ and $N_1=N, N_2=1$ are trivially the ELS model, which has the critical threshold
$\sigma_c=1/4$, when $N\to\infty$. Because load confinement enhances the accumulation of damage in the sub-units, for all intermediate sizes the critical load falls below this limit. 

\begin{figure*}[tbh]
\centering
\includegraphics[width=12cm, keepaspectratio]{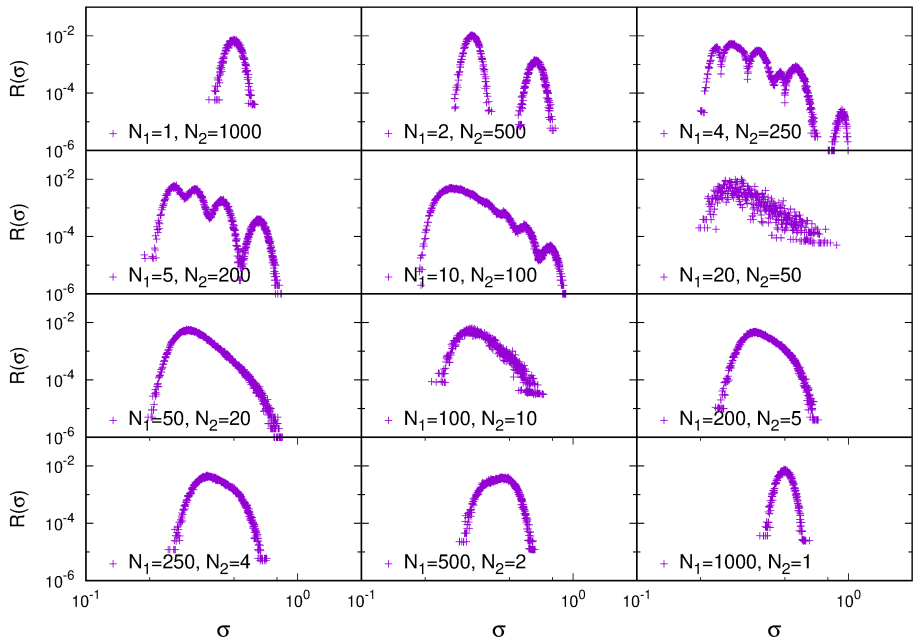}
\caption{Distribution of loads on elementary fibers in two-level hierarchical fiber bundles at the point of failure, system size $N=10000$ fixed, distributions are averaged over an ensemble of fiber bundle realizations. The width of the distributions is largest for intermediate $N_1, N_2$ values, where the system is weak.}
\label{two_level_stresses}
\end{figure*}

For illustration purposes, let us consider the case $N_1=2, N_2=N/2$. This is indeed equivalent to a system of $N/2$ fibers in ELS setting. But the failure threshold distributions of those $N/2$ fibers are different from the uniform distribution. To calculate the failure threshold of the whole system, the job really is to calculate the failure threshold distribution of the coupled fibers. Assuming that the lowest level fibers are from a uniform distribution in $(0:1)$, if a load $\sigma$ is applied, the probability that both fibers fail is:
\begin{itemize}
\item for $\sigma\le0.5$
   \begin{itemize}
      \item when both the thresholds are below $\sigma$; probability for that is $\sigma^2$
      \item when one is below $\sigma$ and the other one is $\sigma\ge\sigma_t<2\sigma$; probability for that is $2\sigma(2\sigma-\sigma)$
   \end{itemize}
\item for $\sigma>0.5$
  \begin{itemize}
     \item one of the thresholds is below $\sigma$; probability for that is $2\sigma-\sigma^2$
  \end{itemize}
\end{itemize}
Therefore, considering the total probabilities for the two cases above, the cumulative distribution $P(\sigma)$ of the failure threshold becomes:
\begin{eqnarray}
P(\sigma) &=& 3\sigma^2 \hskip0.2cm \text{for} \hskip0.2cm \sigma\le0.5 \nonumber \\
&=& \sigma(2-\sigma) \hskip0.2cm \text{otherwise,}
\end{eqnarray}
giving the corresponding probability distribution $p(\sigma)$ as
\begin{eqnarray}
p(\sigma)&=& 6\sigma \hskip0.2cm \text{for} \hskip0.2cm \sigma\le 0.5 \nonumber \\
&=& 2-2\sigma \hskip0.2cm \text{for} \sigma>0.5.
\end{eqnarray}
For an ELS fiber bundle where the distribution and the cumulative distribution of the failure thresholds are respectively $p(\sigma)$ and $P(\sigma)$,
the critical elongation per fiber $\Delta_c$ follows \cite{fbm_rmp}
\begin{equation}
1-\Delta_cp(\Delta_c)-P(\Delta_c)=0,
\label{critical_load}
\end{equation}
implying in this case
\begin{equation}
1-9\Delta_c^2=1
\end{equation}
with the only physical solution being $\Delta_c=1/3$. The critical force in turn follows
\begin{equation}
F_c=N(1-P(\Delta_c))\Delta_c,
\end{equation}
giving $\sigma_c=F_c/N=2/9$. Therefore, $\sigma_c^{(2)}=2/9<\sigma_c^{(1)}=1/4$, in the limit $N\to\infty$.
This estimate agrees well with the simulation data shown in Fig. \ref{level2_load}. As mentioned before, this apparently counter-intuitive feature is the result of enhanced damage accumulation due to load confinement within a single unit. In Fig. \ref{two_level_stresses}, the load distribution on the fibers in the last stable configuration before failure (averaged over an ensemble of $\sim 2000$ realizations) is shown for various values of $N_1, N_2$. As can be seen, the distribution function is widest, when the failure threshold is lowest, which implies that some fibers carry much larger loads than the average. Such load concentration leads to enhanced damage accumulation. On either side of this, the load concentration is decreasing due to (1) lack of surviving fibers within one unit (small $N_1$ limit) and (2) due to wider spreading of the load within one unit (large $N_1$ limit). 

\subsubsection*{Avalanche size distribution}
\begin{figure*}[tbh]
\centering
\includegraphics[width=13cm, keepaspectratio]{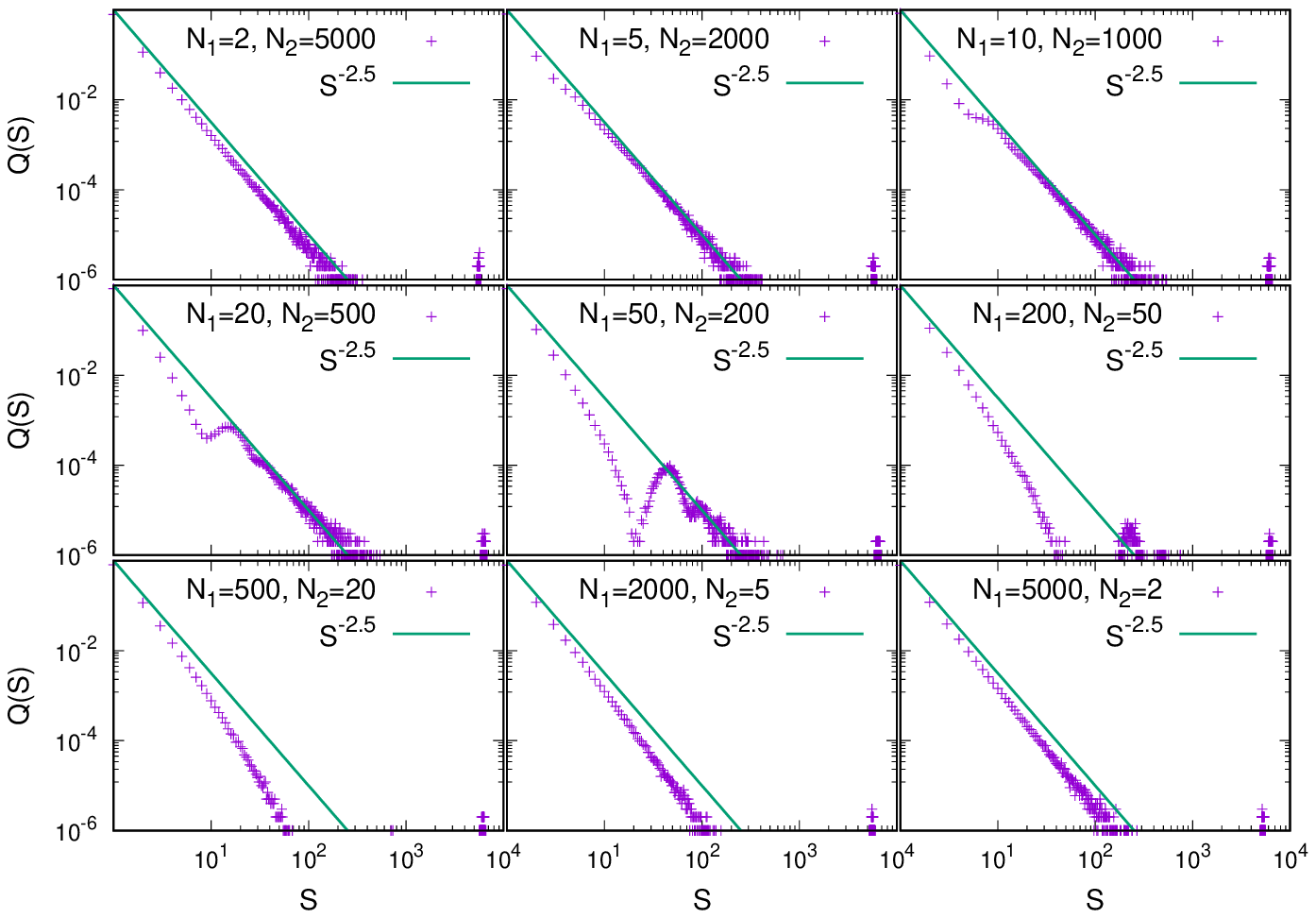} 
\caption{Size distribution of avalanches for the same systems as in Figure \ref{two_level_stresses}; large avalanches that span multiple units show a power law decay with exponent value $-3/2$, smaller avalanches which are confined in one unit, show an exponential decay; the cross-over occurs for avalanche sizes approximately equal to the unit size $N_1$.}
\label{two_level_avalanches}
\end{figure*}

The effect of damage accumulation is also manifested in the avalanche size distribution of the model. Fig. \ref{two_level_avalanches} shows the avalanche size distributions for various values of $N_1, N_2$ with $N=10000$ fixed. Again distributions are averaged over an ensemble of realizations. The generic feature of the avalanche size distribution is to have an exponential drop for smaller avalanches, a power-law tail with an exponent value $-5/2$ at large avalanches, and a crossover that may correspond to a maximum of the probability density function. The initial drop in the distribution is due to the events that are confined within one unit. In that unit, due to stress  localization, the avalanches are exponentially distributed. However, for the large events that spread beyond one unit, the smaller scale of the single unit size is no longer relevant and the avalanche statistics behaves similar to the ELS version of the model, matching its analytically calculable exponent value $-5/2$ \cite{fbm_rmp}. 

Obviously, in the limiting cases with $N_1=1$ and $N_1=N$, the avalanche size distribution follows the power-law 
function for the full range. In all intermediate cases, a signature of the hierarchical structure is manifested in the avalanche size distribution by a cross-over from exponential to power law behavior, which occurs at an avalanche size $S \approx N_1$ that approximately equals the unit size. 

\subsection{Three level hierarchy} 

As the next step of generalization, we study the three level hierarchical fiber bundle model with $N_3$ fibers
at the highest level, each of which consists of $N_2$ level-two fibers, of which each is made up of $N_1$ fibers at the lowest level. As before, we keep $N=N_1N_2N_3$ to be fixed. 

The critical load of the system, as before, is affected by the hierarchical structure. As an example, one can calculate
the critical load for the case $N_1=2, N_2=2, N_3=N/4$. Assuming that at the lowest level, the failure threshold distributions 
of the fibers are drawn from a uniform distribution in $(0:1)$, the probability that one element at the highest level (consisting 
of two elements in the intermediate level, which in turn are made up of two fibers at the lowest level) fails when a load $\sigma$
is applied, is given by (for $\sigma\le 0.5$)
\begin{itemize}
 \item when both the elements in the intermediate level have failure threshold below $\sigma$; probability of that: $\left[\int\limits_0^{\sigma}6\sigma d\sigma\right]^2=9\sigma^4$. 
\item Probability that one element in the intermediate level has failure threshold below $\sigma$, and the other element has failure threshold 
between $\sigma$ and $2\sigma$ is: $2\left[\int\limits_0^{\sigma}6\sigma d\sigma\right]\left[\int\limits_{\sigma}^{2\sigma}6\sigma d\sigma\right]=27\sigma^4$. 
\end{itemize} 
Consequently, the cumulative distribution and the probability distribution for the failure thresholds at the highest level of hierarchy
 respectively becomes (for $\sigma\le 0.5$):
$P(\sigma)=63\sigma^4$ and $p(\sigma)=252\sigma^3$. 
The critical load follows Eq. (\ref{critical_load}), giving
$\Delta_c=\sqrt[\leftroot{-3}\uproot{3}4]{\tfrac{1}{315}}$ and $\sigma_c=(1-\frac{63}{315})\sqrt[\leftroot{-3}\uproot{3}4]{\tfrac{1}{315}} \approx 0.1898$.
Comparing with simulations, for $N_1=2, N_2=2, N_3=2500$, $\sigma_c \approx 0.1915$ and $N_1=2, N_2=2, N_3=5000$, $\sigma_c\approx 0.1912$, the trend is
towards the analytical estimate as $N$ increases. 

For the total range of $N_1, N_2, N_3=N/(N_1N_2)$,  the critical loads tend to the ELS limit $1/4$ if either $N_1$, $N_2$, or $N_3$ becomes comparable to $N$. The load bearing capacity is smaller for all the intermediate configurations where the effect of confinement-enhanced damage accumulation is most prominent. Notably, in all hierarchical configurations, the critical load is lower than for the ELS model.  
\begin{figure*}[tbh]
\centering
\includegraphics[width=13cm, keepaspectratio]{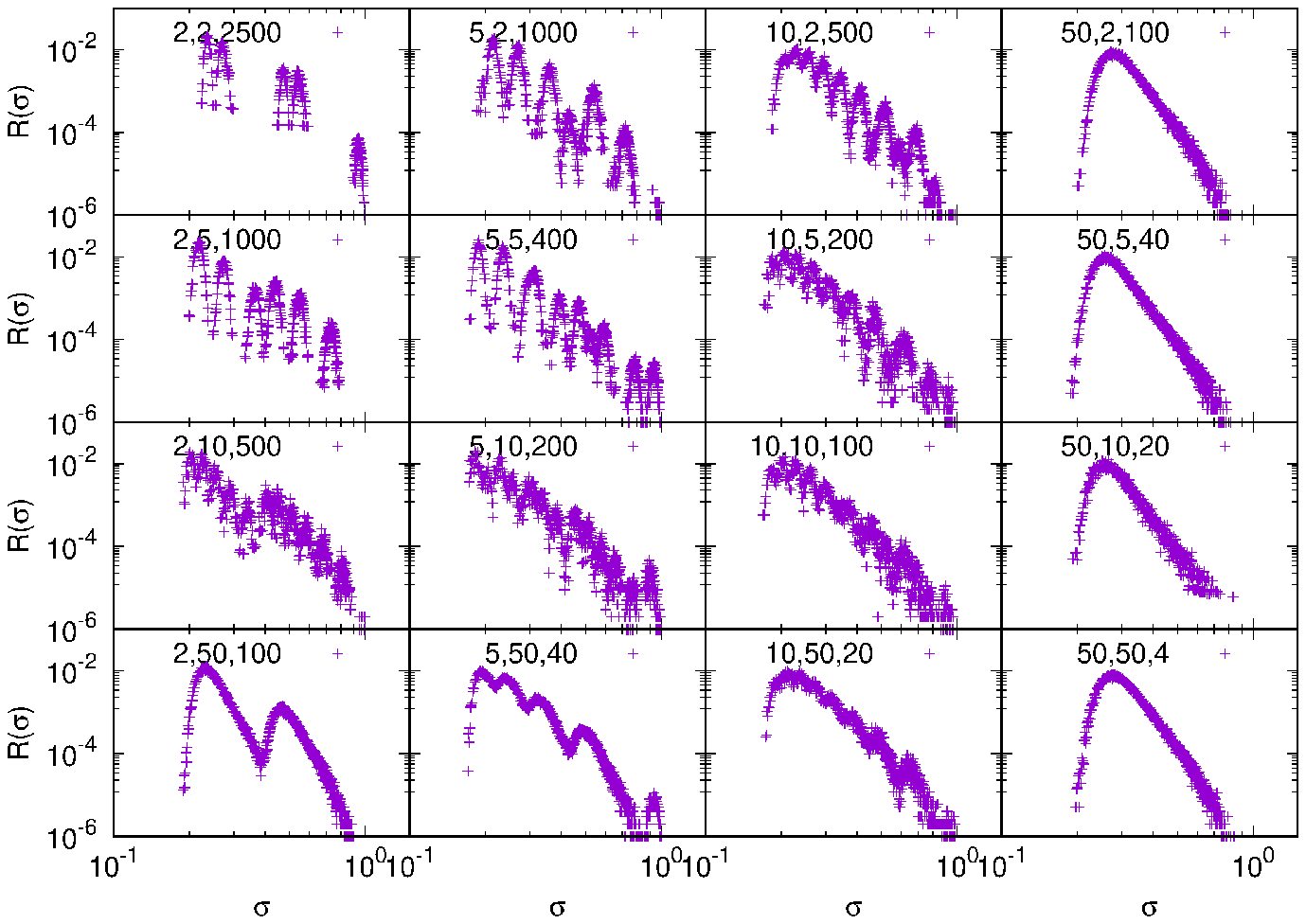} 
\caption{Distribution of loads on elementary fibers in three-level hierarchical fiber bundle models at the point of failure, system size $N =10000$ fixed, for different values of $N_1, N_2$ and $N_3=N/(N_1N_2)$. The effects of stress confinement is evident for intermediate values of $N_1, N_2$, where the stress distribution is widest.}
\label{three_level_stresses}
\end{figure*} 

\begin{figure*}[tbh]
\centering
\includegraphics[width=13cm, keepaspectratio]{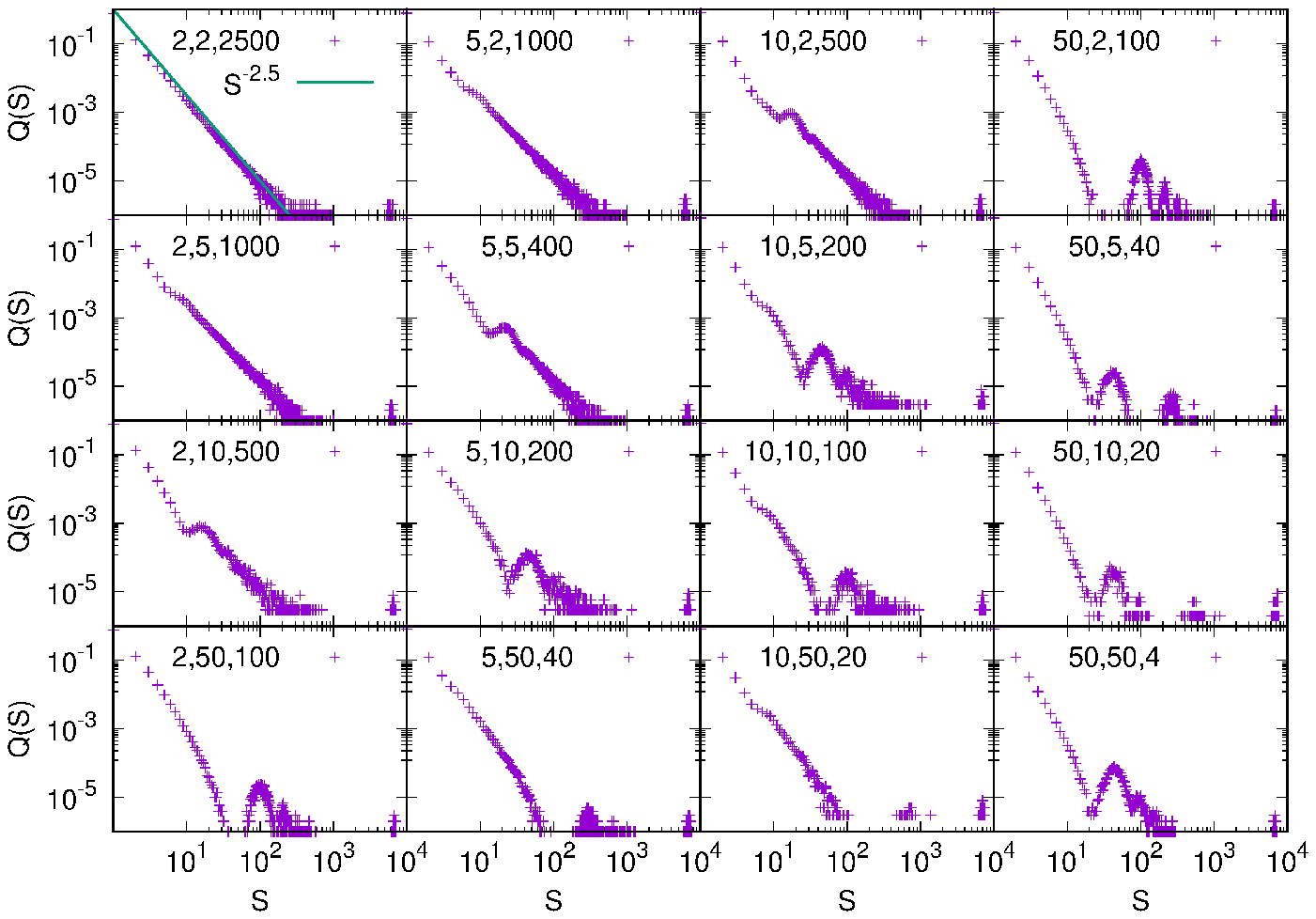} 
\caption{Size distribution of avalanches for the same systems as in Figure \ref{three_level_stresses}; large avalanches that span multiple units show a power law decay with exponent value $-3/2$, smaller avalanches which are confined in units of lower hierarchy show an exponential decay.}
\label{three_level_avalanches}
\end{figure*} 

Fig. \ref{three_level_stresses} shows the distribution of the load per lowest-order fiber in the last stable configuration. The width of the distribution is maximum for the lowest failure thresholds, confirming the effect of confinement-enhanced damage nucleation.

\subsubsection{Avalanche size distribution}
The avalanche size distribution, shown in Fig. \ref{three_level_avalanches}, reflects the damage nucleation process described above. As before, large avalanches which span many units of the lowest and/or the intermediate hierarchy level, define a $-5/2$ power-law tail. But the effect of damage nucleation is visible in smaller avalanches that do not span multiple units and are exponentially decaying in size. There are two length scales, corresponding to the two levels of lower hierarchies, where such decays are seen. 

\subsection{Higher level hierarchy}

In cases where the number of levels is high, but the size of each level is small, the effect of damage accumulation is rather intriguing. As shown before, there will be a length-scale cut-off for each level, but the number of such cut-offs is high. The overall distribution, therefore, is affected at multiple scales. 
\begin{figure}[tbh]
\centering
\includegraphics[width=8cm, keepaspectratio]{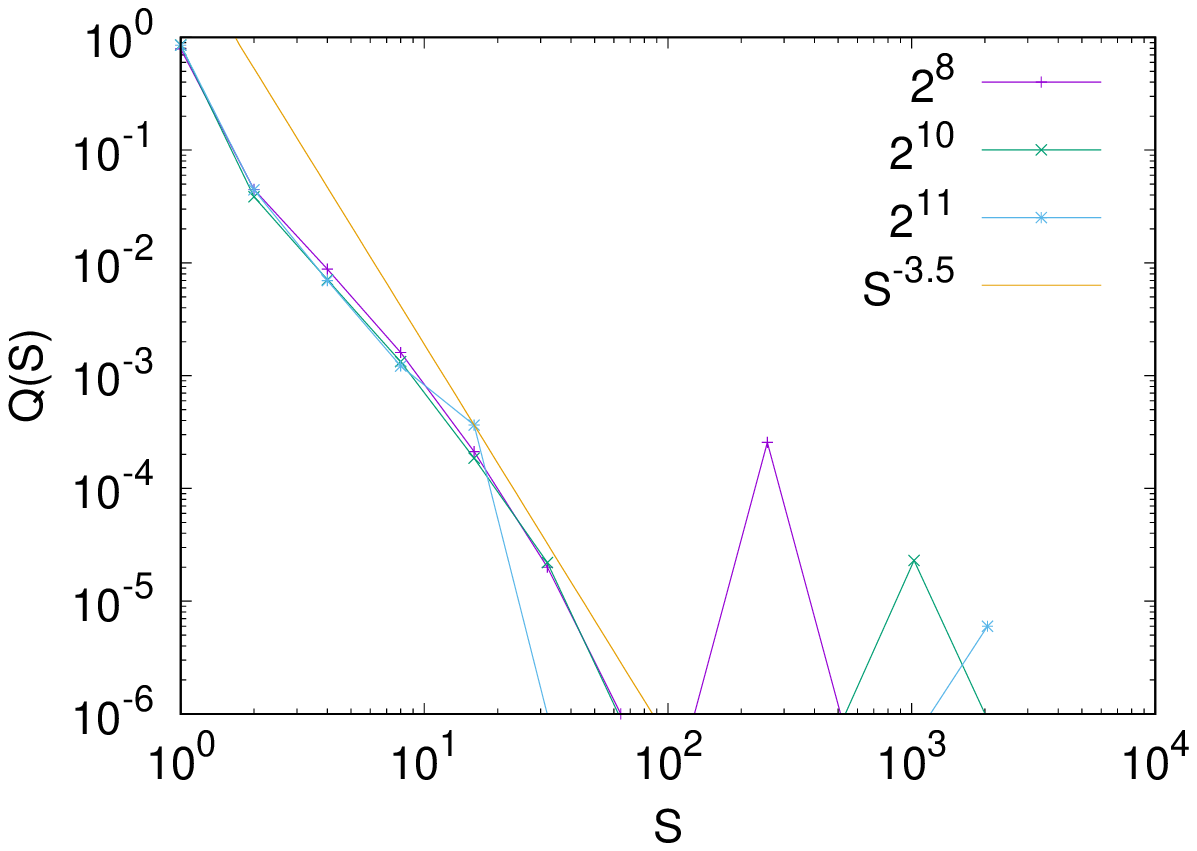} 
\caption{Size distributions of avalanches for a hierarchical structure with 2 fibers in each unit and multiple levels of hierarchy are shown.}
\label{high_hier}
\end{figure} 
Due to the exponential growth in the system size with the number of levels, only a limited range is accessible for numerical analysis. In analyzing this case, we use the same value $N_i = N_1 =2$ value for all hierarchy levels, producing a self similar structure. Fig. \ref{high_hier} shows avalanche size distributions for various values of the number of levels. The avalanche size distributions have power-law characteristics but are much steeper than what is expected for a ELS reference structure. The simulations are shown for a Weibull distribution, the results are also similar for the uniform distribution of thresholds. 

\subsection{Maximizing the global failure threshold}

A common observation for the hierarchical fiber bundle models studied here is that the failure threshold of any hierarchical structure is lower than the failure threshold of a system with equal size and one level of hierarchy (i.e. the usual FBM). In this section, we deal with the question whether the failure threshold of the hierarchical structure could be increased beyond the limit of the single level arrangement. 

The idea is to use a set of elements with failure thresholds drawn from a given distribution, to classify them according to their failure threshold, and then to arrange them in a suitable hierarchical structure to maximize the effective failure threshold of the system. Such situations could arise in various design problems with shared load, such as the power grid network, computer redundancies, traffic networks and so on. The crucial feature is to group the elements into units by correlating the unit size with the failure thresholds of the individual elements. (Of course, in practice this depends on availability of a method to determine the failure threshold of a unit without destroying it). For simplicity, we consider
a two level hierarchy. 

A fiber bundle is strongest if all fibers have exactly the same strength. In a two-level hierarchical structure, one can try to exploit this observation at the lower hierarchical level by a suitable grouping of fibers which collects level-one fibers of approximately equal strength into the same level-two fiber. At the higher level, one needs again to ensure that all the resulting level-2 fibers have approximately equal strength. A straightforward method to achieve these dual objectives is the following:

Let the elementary fibers be arranged in the ascending order of their failure thresholds. Then divide this ordered list into $m$ groups with population fractions $n_i$, with $\sum\limits_i^mn_i=1$.  If the average value of the failure thresholds in each group is $f_i$, then 
\begin{equation}
\sum\limits_{i=1}^mn_if_i=\frac{1}{2}
\label{threshold_sum}
\end{equation}
for a uniform distribution in $(0:1)$. We now require that the numbers and failure thresholds in each group fulfill the relation
\begin{equation}
\frac{n_i}{n_{i+1}}=\frac{f_{i+1}}{f_i}.
\label{threshold_ratio}
\end{equation}
Combining Eq. (\ref{threshold_sum}) and Eq. (\ref{threshold_ratio}), we can see that this requirement is tantamount to 
\begin{equation}
n_if_i=\frac{1}{2m},
\label{2m_limit}
\end{equation}
i.e., each group has the same average load bearing capacity.

Now, if the partition between $i-1$st  and $i$th group in terms of threshold values is at $x_i$, then 
\begin{equation}
n_i=x_i-x_{i-1},
\label{n_equation}
\end{equation}
and 
\begin{equation}
f_i=x_{i-1}+\frac{x_i-x_{i-1}}{2}.
\label{f_equation}
\end{equation}
Putting Eq. (\ref{n_equation}) and Eq. (\ref{f_equation}) into Eq. (\ref{2m_limit}), we get
\begin{equation}
(x_i-x_{i-1})\frac{(x_i-x_{i-1})}{2}=\frac{1}{2m},
\end{equation}
or
\begin{equation}
x_i^2-x_{i-1}^2=\frac{1}{m} \hskip0.2cm \text{with} \hskip0.2cm x_m=1, x_0=0.
\end{equation}
This implies $x_i=\sqrt{\frac{i}{m}}$. 

Next, we make use of the fact that, in a ELS fiber bundle, the weakest fibers may well be rather useless in terms of the strength of the system. This can be easily seen by looking at a ELS fiber bundle with $N$ fibers and uniformly distributed thresholds on the interval $(0:1)$, that has a critical stress of $\sigma_c = 0.25$ and thus carries a load of $0.25 N$: Throw away the weaker half of the fibers to create a bundle with $N/2$ fibers and thresholds on the interval $(0.5:1)$, which has a critical stress of $0.5$ and carries a load of $0.25 N$ still. This simply follows from the fact that at the critical point, the weaker half of the system is broken and it is the stronger half that carries the total load.  We apply a similar strategy by lumping, prior to partitioning, all level-one fibers with a threshold below $\sigma$ into a single level-two fiber $i=0$ which means that in the thermodynamic limit $N,m \to \infty$ they become irrelevant to the dynamics of the system. 

\begin{figure}[tbh]
\centering
\includegraphics[width=8cm, keepaspectratio]{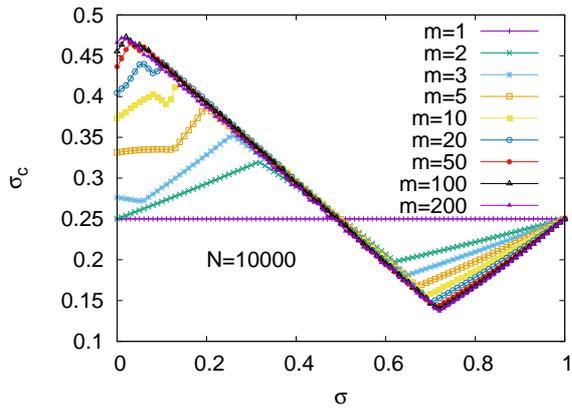} 
\caption{The critical load $\sigma_c$, as a function of the partitioning parameter $\sigma$ in Eq. (\ref{partition}). The maximum strength is obtained for large $m$ values when the groups are well partitioned into almost equal strengths, and is substantially higher than the single hierarchy limit $1/4$, which is in turn higher than all other hierarchical arrangements.}
\label{hier_max}
\end{figure}

The above calculation can, then, be repeated for the remaining fraction $(1-\sigma)$ of fibers, with new variables $x_i^{\prime}$, which are related to the old variables $x_i$ as
\begin{equation}
x_i^{\prime}=\frac{x_i-\sigma}{1-\sigma}.
\end{equation}
Therefore, the optimized partitioning of the fibers in hierarchical modules under a threshold $\sigma$ is
\begin{equation}
x_i=\sqrt{\frac{i}{m}}(1-\sigma)+\sigma,
\label{partition}
\end{equation}
when there are $m$ units in the second level of hierarchy, in a two-level hierarchical system. Finally, to answer the question for what value of $\sigma$ an optimal configuration may be obtained,  Fig. \ref{hier_max} shows the value of the critical load $\sigma_c$ for two-level hierarchical fiber bundles partitioned according to Eq. (\ref{partition}) with various values of the truncation parameter $\sigma$ and the number of groups $m$. For $m=1$, the critical load is trivially at $0.25$. For all other values of $m$, there is at least some values of $\sigma$ for which the critical load is higher than this reference value. Indeed, the gain of minimizing the effects of weaker fibers, as described above, could be seen for small $\sigma$ values. On the other hand, for large $m$, the system gets well partitioned into $m$ equally strong groups. Under that circumstance, given that the average threshold of the whole system cannot change, the only option is that the strength of each group tend to that average value i.e. $0.5$. Therefore we see that the critical strength tends to $0.5$ for high $m$ at low $\sigma$ values. This implies a $100\%$ increase in the strength, and is much higher than other attempts to maximize strength \cite{biswas_sen}. 

Therefore, we find that for a suitable arrangement of the hierarchical structure, it is possible to reach close to the maximum possible failure threshold under uniform loading and uniform load redistribution.

\section{Discussion and conclusions} 
 
The hierarchical nature of a heterogeneous material can have major consequences on its dynamics and load bearing capacity. However, we consistently find that, for fiber bundle models, a hierarchical arrangement generally reduces the global failure threshold as opposed to the one without such an arrangement. The reason behind the lower failure threshold for hierarchical fiber bundle structures is simply the mean-field nature of the fiber bundle model. Due to the mean-field nature, the broadest possible range of load redistribution is the one where all fibers can equally share the load of a failed fiber. For any hierarchical arrangement, there is necessarily a load localization, since the load of a failed fiber must be shared by fibers belonging to the same hierarchical unit. This will always make a part of the system inaccessible for load redistribution following the failure of a given fiber, and the local confinement of load enhances local damage accumulation. This effect of load or damage localization due to hierarchical structures is also manifest in the avalanche dynamics of the system. In the avalanche dynamics, large avalanches that span several hierarchical units, behave like in the mean field model. However, smaller avalanches, that are confined within single units, differ significantly from the mean-field limit. Indeed, looking at the avalanche size distributions, is a possible way to understand the hierarchical nature of a material, for which the detailed internal structures are not known.

However, we have also demonstrated that the confinement of load redistribution within a unit, following a local failure in that
unit, could be used to enhance the load bearing capacity of the entire system. The key idea is to make all units having same 
total capacity i.e. combining large number of weaker fibers in one group and small number of stronger fibers in another and 
intermediate sizes in between. A specific recipe of such a division is proposed, which can significantly increase the 
total capacity. In practice this is contingent upon knowledge of the fiber failure thresholds, which depends on availability of a suitable non-destructive testing method.

In conclusion, we have studied the avalanche dynamics of hierarchical fiber bundles and have shown that its nature bears
the signature of the underlying hierarchy. While the failure thresholds of the fiber bundles in usual hierarchical structures 
are less than the one without such structures, we propose a mechanism in which the critical load could be made much higher. The observation that hierarchy does not necessarily make a structure stronger matches findings on the failure strength of hierarchical fuse models, where hierarchically organized structures were found to be slightly weaker than non hierarchical reference structures \cite{scirep}. On the other hand, the main advantage of hierarchical structures, namely that they effectively suppress crack propagation driven by crack-tip stress concentrations and therefore possess a high degree of flaw tolerance \cite{scirep}, has no counterpart in fiber bundle models which by their nature are devoid of spatial structure.  

\begin{acknowledgments}
We acknowledge funding by DFG under grant No. Za 171-9. 
\end{acknowledgments}


\end{document}